\documentclass[conference]{IEEEtran}
\IEEEoverridecommandlockouts
\usepackage{cite}
\usepackage{amsmath,amssymb,amsfonts}
\usepackage{algorithmic}
\usepackage{graphicx}
\usepackage{textcomp}
\usepackage{xcolor}

\usepackage{hyperref}

\usepackage{caption}
\usepackage{subcaption}
\usepackage{listings}

\definecolor{red}{rgb}{1,0.,0}
\definecolor{comment-text-color}{rgb}{0,0.9,0.1}
\usepackage{tikz}
\usetikzlibrary{quantikz}
\usetikzlibrary{backgrounds,fit,decorations.pathreplacing,calc}

\def\BibTeX{{\rm B\kern-.05em{\sc i\kern-.025em b}\kern-.08em
    T\kern-.1667em\lower.7ex\hbox{E}\kern-.125emX}}
    
\begin{document}

\title{\textit{Really} Embedding Domain-Specific Languages into C++}
\makeatletter
\newcommand{\linebreakand}{%
  \end{@IEEEauthorhalign}
  \hfill\mbox{}\par
  \mbox{}\hfill\begin{@IEEEauthorhalign}
}
\makeatother
\author{
\IEEEauthorblockN{Hal Finkel}
\IEEEauthorblockA{\textit{Leadership Computing Facility} \\
\textit{Argonne National Laboratory}\\
Lemont, IL, USA \\
hfinkel@anl.gov \thanks{This manuscript has been authored by UT-Battelle, LLC under Contract No. DE-AC05-00OR22725 with the U.S. Department of Energy. The United States Government retains and the publisher, by accepting the article for publication, acknowledges that the United States Government retains a non-exclusive, paid-up, irrevocable, world-wide license to publish or reproduce the published form of this manuscript, or allow others to do so, for United States Government purposes. The Department of Energy will provide public access to these results of federally sponsored research in accordance with the DOE Public Access Plan. (http://energy.gov/downloads/doe-public-access-plan).}}
\and
\IEEEauthorblockN{Alexander McCaskey}
\IEEEauthorblockA{ \textit{Computer Science and Mathematics Division} \\
\textit{Oak Ridge National Laboratory} \\
Oak Ridge, TN, USA \\
mccaskeyaj@ornl.gov}
\and
\IEEEauthorblockN{Tobi Popoola}
\IEEEauthorblockA{\textit{Boise State University}\\
Boise, ID, USA \\
tobipopoola@u.boisestate.edu}
\linebreakand
\IEEEauthorblockN{Dmitry Lyakh}
\IEEEauthorblockA{\textit{Leadership Computing Facility} \\
\textit{Oak Ridge National Laboratory} \\
Oak Ridge, TN, USA \\
liakhdi@ornl.gov}
\and
\IEEEauthorblockN{Johannes Doerfert}
\IEEEauthorblockA{\textit{Leadership Computing Facility} \\
\textit{Argonne National Laboratory}\\
Lemont, IL, USA \\
jdoerfert@anl.gov}
}

\maketitle

\begin{abstract}
Domain-specific languages (DSLs) are both pervasive and powerful, but remain difficult to integrate into large projects. As a result, while DSLs can bring distinct advantages in performance, reliability, and maintainability, their use often involves trading off other good software-engineering practices. In this paper, we describe an extension to the Clang C++ compiler to support syntax plugins, and we demonstrate how this mechanism allows making use of DSLs inside of a C++ code base without needing to separate the DSL source code from the surrounding C++ code. 
\end{abstract}


\section{Introduction}
Domain-specific languages play an important part in our modern software ecosystem. Domain-specific languages are commonly used to generate lexical analyzers (e.g., Lex~\cite{lesk1975lex}, re2c~\cite{bumbulis1993re2c}) and parsers (e.g., Yacc~\cite{johnson1975yacc}, ANTLR~\cite{parr1995antlr}). In high-performance computing, DSLs are used to generate numerical kernels (e.g., SPIRAL~\cite{franchetti2018spiral}, TCE~\cite{hirata2003tensor}, LGen~\cite{spampinato2014basic}, Linnea~\cite{barthels2020automatic}, TACO~\cite{kjolstad2017tensor}, Devito~\cite{lange2016devito}), and many code bases use custom code generators (e.g., the codelet generator in FFTW~\cite{frigo1998fftw}, Kranc in the Einstein Toolkit~\cite{husa2006kranc}).

Unfortunately, these DSLs are often difficult to integrate into a larger software project. At its root, one significant source of difficulty is due to the need to provide the input to the DSL's translator in one or more source files separate from the source files containing the bulk of the project's source code. The input to the DSL, however, needs to be somehow integrated with the code is the rest of the project. The DSL-generated code provides and/or uses interfaces provided by the rest of the source code, sometimes making use of non-trivial types defined elsewhere in the project. Some DSL's (e.g., Lex, Yacc) allow the programmer to embed code in a host, general-purpose programming language (e.g., C/C++) into the DSL's input source files. While relatively convenient, and often necessary for performance, this increases the number of, and complexity of, dependencies between the DSL source and the rest of the project. Practically, integration of DSLs into project build systems requires effort, and sometimes, specialized skills. Most importantly, programmers often try to keep related functions and/or components together in the same source file to make the source code easier to understand. While being forced to keep the DSL's input in files separate from the rest of the project source code may seem to be a trivial inconvenience, it's not. A majority, 60-90\%, of software-development costs are associated with reading and navigation source code as part of maintenance tasks~\cite{erlikh2000leveraging}. Consistent with our experience, when examining how programmers use their editors to perform these kinds of tasks, 35\% of their time, on average, is spent navigating between dependencies~\cite{ko2005eliciting}. Thus, the productivity of using DSLs can be significantly improved by reducing logically-undesirable partitioning of source code between different files.

In this paper, we describe a new technique for integrating DSLs into C++ source code: the syntax plugin. The implementation is built on the Clang C++ compiler~\cite{lattner2008llvm}. Clang can compile C++ code to executable form using the LLVM compiler infrastructure~\cite{lattner2004llvm}, but can also directly perform source-code analysis~\cite{kremenek2008finding} and rewriting. While performing any of these tasks, Clang can make use of user-provided plugins. However, no existing plugin interface provides the ability to integrate DSLs as described here, and so as described below, a new plugin interface was designed and implemented. While we do not quantitatively evaluate productivity gains, or other factors affecting developer willingness to use DSLs, in this paper, it is our sense that the presented technique increases developer productivity and otherwise lowers barriers to adopting DSLs inside of C++ code bases.

The implementation of Clang with support for syntax plugins is available from: \url{https://github.com/hfinkel/llvm-project-csp}

The remainder of this paper is organized as follows: Section~\ref{sec:cp} reviews Clang's plugin support, Section~\ref{sec:sp} describes the implementation of the syntax-plugin infrastructure within Clang, Section~\ref{sec:taco} describes how syntax plugins enable embedding TACO into C++, Section~\ref{sec:qcor} describes embedding programming languages for quantum computing into C++, Section~\ref{sec:taprol} describes embedding a DSL for tensor computation into C++, and Section~\ref{sec:con} concludes with a discussion of future directions.

\section{Background: Clang Plugins}
\label{sec:cp}

Clang plugins are loaded from shared libraries provided on Clang's command line and are integrated into Clang's source-code processing. Figure~\ref{fig:cmdlplugin} shows the relevant command-line arguments.

\begin{figure}
         \begin{lstlisting}
clang++ -c source.cpp \\
  -fplugin=/path/to/somePlugin.so
         \end{lstlisting}
        \caption{An example showing how Clang can be invoked so that it loads (from a shared library named \texttt{somePluginLib.so}) and uses a plugin.}
        \label{fig:cmdlplugin}
\end{figure}

A Clang plugin takes the form of a set of classes, each of which derives from some appropriate \textit{Handler} class, and each of which is registered using an appropriate static object. Clang provides several kinds of handlers~\cite{ClangPlu29}:
\begin{itemize}
    \item \texttt{PragmaHandler} - Used by a plugin to provide new kinds of pragmas.
    \item \texttt{ParsedAttrInfo} - Used by a plugin to provide new kinds of attributes.
    \item \texttt{PluginASTAction} - Used by a plugin to provide an \textit{AST listener}, an object that can observe AST-node creation events.
\end{itemize}

As illustrated in Figure~\ref{fig:pluginex}, a single plugin can have multiple handlers, including multiple kinds of handlers, and use all of them together in order to provide its functionality.

\begin{figure}
         \begin{lstlisting}
#include "clang/Frontend/FrontendPluginRegistry.h"
... // other includes.
using namespace clang;

namespace {
class ExampleASTConsumer : public ASTConsumer {
  CompilerInstance &Instance;

public:
  ExampleASTConsumer(CompilerInstance &Instance)
      : Instance(Instance) {}

  bool HandleTopLevelDecl(DeclGroupRef DG)
    override {
    // Do something to handle a new declaration.
    return true;
  }
  
  void HandleTranslationUnit(ASTContext& context)
    override {
    // Do something to handle the completion of
    // the translation unit.
  }
};

class ExampleASTAction : public PluginASTAction {
protected:
  std::unique_ptr<ASTConsumer> CreateASTConsumer(
    CompilerInstance &CI, llvm::StringRef)
    override {
    return
      std::make_unique<PrintFunctionsConsumer>(CI);
  }
  
  bool ParseArgs(
    const CompilerInstance &CI,
    const std::vector<std::string> &args)
    override {
    // Handle custom plugin command-line arguments.
    return true;
  }
  
  void PrintHelp(llvm::raw_ostream& ros) {
    ros << "A help message goes here\n";
  }
};

class ExamplePragmaHandler : public PragmaHandler {
public:
  ExamplePragmaHandler() :
    PragmaHandler("an_example") { }

  void HandlePragma(Preprocessor &PP,
    PragmaIntroducer Introducer,
    Token &PragmaTok) override {
    // Handle an encountered:
    //   #pragma an_example more tokens
  }
} // anonymous namespace

static FrontendPluginRegistry::Add<ExampleASTAction>
X("example-plugin", "an example sketch");

static PragmaHandlerRegistry::
  Add<ExamplePragmaHandler>
Y("example-plugin-cntrl","enable something");
         \end{lstlisting}
        \caption{An example showing how Clang plugins are structured. This example sketches a plugin using the \texttt{PragmaHandler} interface to provide a new pragma and the \texttt{PluginASTAction} to observe relevant AST-node creation events.}
        \label{fig:pluginex}
\end{figure}

It is worth noting that this interface is tied directly to Clang's AST data structures. This means that the plugins are not only specific to Clang, but also, in practice, tied to the particular version of Clang against which the plugin was compiled. Programming the plugin requires knowledge of the data structures used by Clang's lexical analysis and parsing infrastructure, along with Clang's AST. For a more-complete example, we refer the reader to \texttt{examples/AnnotateFunctions} in Clang's source repository~\cite{clangannotex}.

\section{Clang Syntax Plugins}
\label{sec:sp}

None of the existing plugin interfaces allow the plugin to alter the fundamental syntax accepted by Clang's parser. Here we explore a plugin interface that allows for exactly that: within a specifically-tagged function, the function body may contain code that is not valid C++ code. The code in the function body must still obey certain rules, but only a few:
\begin{itemize}
    \item The code must admit a valid C++ tokenization. Those tokens might not form a valid fragment of C++ code, but the code must be describable as C++ tokens.
    \item The code, like all of the token stream, is subject to C++ macro expansion. That must be acceptable for the plugin's use cases.
    \item The code must use balanced C++ delimiters - the implementation must be able to find the '\}' character matching the \{ starting the function body, and it does so using C++ token-capturing rules, without consulting the plugin.
\end{itemize}

The task of the plugin, when handling an appropriately-tagged function, is to generate a replacement stream of text containing a valid fragment of C++ code. This text stream effectively \textit{replaces} the tagged function definition. Thus, the DSL takes its input and transforms it into some other C++ code that is processed by Clang in the usual way. The whole interface, illustrated in Figure~\ref{fig:shex} is fairly simple, consisting of only two functions:
\begin{itemize}
    \item \texttt{GetReplacement} - Takes the token stream and provides the replacement stream of C++ code.
    \item \texttt{AddToPredefines} - Adds to the stream of C++ code parsed at the beginning of the translation unit. Useful for adding \#include directives and declarations required by any replacement code.
\end{itemize}

\begin{figure}
         \begin{lstlisting}
#include "clang/Frontend/FrontendPluginRegistry.h"
... // other includes.
using namespace clang;

namespace {
class ExampleSyntaxHandler : public SyntaxHandler {
public:
  ExampleSyntaxHandler() :
    SyntaxHandler("example") { }

  void GetReplacement(Preprocessor &PP,
                      Declarator &D,
                      CachedTokens &Toks,
                      llvm::raw_string_ostream &OS)
    override {
    // The plugin might handle the token stream
    // directly. It might convert it back into a
    // string. This is common when interfacing
    // with some pre-existing tool or library
    // that expects a source string or file.
    std::string All;
    for (auto &Tok : Toks)
      All += PP.getSpelling(Tok);
      
    ... // Write any local declarations or
        // definitions needed by the
        // replacement code.

    OS << getDeclText(PP,D) << "{\n";
    ... // Write the replacement function body.
    OS <<"}\n";
  }

  void AddToPredefines(llvm::raw_string_ostream &OS)
    override {
    OS << "#include <something.h>\n";
    ... // other definitions
  }
};

} // anonymous namespace

static SyntaxHandlerRegistry::
Add<ExampleSyntaxHandler>
X("syn-example", "example syntax handler");
         \end{lstlisting}
        \caption{An example showing a Clang syntax handler.}
        \label{fig:shex}
\end{figure}

The syntax handler registers the name of the syntax, and the implementation looks for functions with the C++ attribute \texttt{[[clang::syntax(\textit{name})]]}. After the function declaration is parsed, the tokens forming the function body are collected, translated by the plugin, and then the replacement stream is injected into Clang's input in much the same way as an include file is processed.

It was discovered that in most non-trivial use cases, the replacement code in the function body required the declaration or definition of other types and functions. Not everything could appear strictly within the body of the function. This presented an implementation challenge: given that the original function declaration has already been parsed and added to the AST by the time the replacement text is generated, and that Clang's AST design makes it difficult to remove the function declaration that has already been processed, how can new declarations and definitions be injected prior to the replaced function body? Strictly speaking, this is not something Clang's AST supports. However, we were able to elide the ordering problem with the following technique:
\begin{itemize}
    \item The original function declaration is completed with a body containing only a call to \texttt{\_\_builtin\_unreachable()}.
    \item The original function declaration cannot be removed, but it can be renamed, and name in the original function declaration is prefixed with '\_\_' and a non-conflicting string.
    \item A utility function, \texttt{getDeclText} is provided so that the plugin can easily regenerate the original function declaration after providing whatever other definitions and declarations should proceed it.
\end{itemize}

Note that the plugin's \texttt{GetReplacement} function is passed the function's \texttt{Declarator} object. This provides access to the function name, type, parameters, and other properties. It is often the case that knowing the names and types of the function parameters makes it easier to support the direct use of those parameters in the DSL source code.

The next few sections provide examples of syntax plugins that we have developed, and show how they can provide convenient integration of different kinds of DSLs into C++.

\section{TacoPlug: Embedding Linear Algebra in C/C++}
\label{sec:taco}

Linear algebra, and specifically linear algebra with sparse inputs, is a fundamental class of algorithms used in scientific computing~\cite{asanovic2006landscape}. Simple expressions involving matricies or tensors, in order to enable efficient execution on modern, parallel architectures with complex memory hierarchies, must be implemented using a complex series of loop nests. Moreover, the best representation for a given sparse matrix depends on its sparsity pattern and the operations one must perform on it. These factors make linear algebra a good match for DSLs, and motivated the creation of TACO, the Tensor Algebra Compiler~\cite{kjolstad2017tensor}. TACO transforms tensor algebra expressions in index notation, along with information on the desired representations of the tensors, into efficient C/C++ code.

The relative simplicity of the input to TACO also makes it an excellent candidate for embedding into C++. A function computing a expression on some input matrices or tensors can have its body given directly in the natural mathematical notation. Specifically, we have:

\begin{itemize}
	\item Direct support for tensor index-notation syntax in C/C++.
	\item Tensors kernels that can be expressed in different formats.
	\item Support for custom data structures thereby providing easy integration with existing C/C++ applications.
\end{itemize}

\begin{figure}
         \centering
      \begin{subfigure}[b]{\linewidth}
         \centering         
         \begin{lstlisting}
// Custom data structure for some struct <name>
typedef struct <name> {
    ....
    // Plugin requires to and from conversion
    // routines to use custom data structures 
    // with TACOPlug.
    taco_tensor_t* (* <name>2taco)(struct <name> *);
    void (* taco2<name>)
    (taco_tensor_t*,struct <name>* );
} <name>;
    \end{lstlisting}
    \caption{Code illustrating support for custom tensor data structures.}
    \label{fig:tp_cds}
    \end{subfigure}

      \begin{subfigure}[b]{\linewidth}
         \centering         
         \begin{lstlisting}
[[clang::syntax(taco)]] 
    void matrix_vector_mul
    (vector *y,csr *A,vector *x,
         std::string format=
         "  -f=A:ds:0,1 -f=x:d -f=y:d") {
   y(i) = A(i,j) * x(j)
}
    \end{lstlisting}
    \caption{Code illustrating how programmers use TACOPlug to define a function.}
    \label{fig:tp_decl}
    \end{subfigure}

      \begin{subfigure}[b]{\linewidth}
         \centering         
         \begin{lstlisting}
matrix_vector_mul(&vector_y, &matrix_csr_A,
    &vector_x);

    \end{lstlisting}
    \caption{Code illustrating how programmers call the function processed by TACOPlug.}
    \label{fig:tp_call}
    \end{subfigure}
     \hfill
        \caption{Example TACOPlug Usage}
        \label{fig:tp_usage}
\end{figure}

\begin{figure}
         \centering
      \begin{subfigure}[b]{\linewidth}
         \centering         
         \begin{lstlisting}
// Generated by TACO:
int __taco_comput_1(taco_tensor_t *,
    taco_tensor_t *,taco_tensor_t *); 
int __taco_assm_1(taco_tensor_t *,
    taco_tensor_t *,taco_tensor_t *); 

// Assembly Code.
int __taco_assm_1
    (taco_tensor_t *y, taco_tensor_t *A,
    taco_tensor_t *x) {
  int y1_dimension = (int)(y->dimensions[0]);
  ....
  y->vals = (uint8_t*)y_vals;
  return 0;
}

// Compute Code.
int __taco_comput_1(taco_tensor_t *y,
    taco_tensor_t *A, taco_tensor_t *x) {
  ....
  #pragma omp parallel for schedule(runtime)
  for (int32_t i = 0; i < A1_dimension; i++) {
    double tjy_val = 0.0;
    for (int32_t jA = A2_pos[i]; 
        jA < A2_pos[(i + 1)]; jA++) {
      int32_t j = A2_crd[jA];
      tjy_val += A_vals[jA] * x_vals[j];
    }
    y_vals[i] = tjy_val;
  }
  return 0;
}
    \end{lstlisting}
    \caption{Example code generated by TACO.}
    \label{fig:tp_cgen}
    \end{subfigure}

      \begin{subfigure}[b]{\linewidth}
         \centering         
         \begin{lstlisting}
void
mat_vec_mul(vector *y, csr *A, vector *x,
            std::string format=
            "-f=A:ds:0,1 -f=y:d -f=x:d"){
taco_tensor_t * __taco_y = y->vector2taco (y);
taco_tensor_t * __taco_A = A->csr2taco (A);
taco_tensor_t * __taco_x = x->vector2taco (x);
__taco_assm_1(__taco_y,__taco_A,__taco_x);  
__taco_comput_1(__taco_y,__taco_A,__taco_x); 
y->taco2vector(__taco_y,y);
__taco_cleanup_taco(__taco_y);
__taco_cleanup_taco(__taco_A);
__taco_cleanup_taco(__taco_x);
}
    \end{lstlisting}
    \caption{Example code generated by TACOPlug.}
    \label{fig:tp_cgenp}
    \end{subfigure}
     \hfill
     
        \caption{Example TACO/TACOPlug-Generated Code}
        \label{fig:tp_allgc}
\end{figure}

\subsection{TACO - Existing Interfaces}
TACO generates kernels for sparse and dense tensor expressions using a mathematical index notation. The tensor index notation used by TACO is a variation of the work done in by Ricci Curbastro and Levi Civita \cite{Ricci1901}. TACO provides a C++ library interface, which might also be considered a kind of \textit{embedded} DSL, to express tensor index notation using operator overloading and templating. Figure~\ref{fig:tensor_index_notation} shows a tensor index notation for a sparse matrix-vector multiplication. The equivalent code using TACO's C++ library interface is shown in Figure~\ref{fig:taco_dsl}. As can be seen, the mathematical notation is more succinct than that required by the library interface.

TACO also provides a command-line tool that consumes a succinct notation that is closely related to the associated mathematical notation. A sparse matrix-vector multiplication is shown in Figure~\ref{fig:tensor_command_line}. In Figure~\ref{fig:tensor_command_line}, the first argument, \texttt{y(i) = A(i,j) * x (j)}, is the actual tensor-algebra index notation, \texttt{-f=A:ds:0,1} describes the tensor format of \texttt{A}; \texttt{ds} describes a CSR tensor, \texttt{0,1} describes the data layout of the tensor. See~\cite{Chou2018}  for a more-detailed description of the input format and the supported options. 

TACO generates assembly code and compute code. The assemble code allocates memory to store the result of the tensor-algebra computation while the compute code performs the actual computation. TACO uses a unique data structure, taco\_tensor\_t, to store the value and coordinate information of the sparse tensor format. The code generated by TACO performs computation directly on this data structure. The compile call on the object \texttt{y} in Figure~\ref{fig:taco_dsl} generates code as required, compiles the code into a shared library, and dynamically loads it with dlopen.

\subsection{TACOPlug Syntax}

TACOPlug provides support for tensor-index notation in C/C++ using TACO. Figure~\ref{fig:tp_decl} shows TACOPlug's function-declaration syntax. The function body is the tensor index notation describing the computation. The tensor names used in the notation must have corresponding function parameters with the same name. The final argument in the function declaration must be a string variable with a default string specifying the format of the computation. It is necessary to describe the format of each tensor because TACO supports code generation for many combinations of different tensor formats. The format specification follows closely the format specification accepted by the TACO command-line tool.

\subsection{TACOPlug Code Generation}
TACOPlug uses the syntax-handler interface to allow the definition of functions computing sparse-matrix using TACO's sucinct syntax. The function declaration body is parsed to get the actual tensor-index notation and the format specification of the computation. TACOPlug performs a check to ensure that there is an actual parameter for each tensor in the notation. After a successful type check, the tensor-index notation and format specification is sent directly to TACO for code generation. TACO generates compute and assembly code which can be seen in Figure~\ref{fig:tp_cgen}. A unique number is appended to each compute and assembly functions generated by TACO to ensure that all declarations are unique.

\subsection{Supporting Custom Data Structures}
TACO generates code that uses taco\_tensor\_t, therefore for TACOPlug to work with functions taking other data structures, those data structures must have associated conversion procedures to and from taco\_tensor\_t. Specifically, there must be to be \textit{to} and \textit{from} function pointers present in the struct. Figure~\ref{fig:tp_cds} shows a sample custom data structure. The asymptotic cost of converting to and from taco\_tensor\_t for sparse and dense tensors is $O(n)$ where $n$ is the mode of the tensor. TACOPlug automatically calls conversion routines for non-taco\_tensor\_t parameter types. Support for custom data structures is essential for integration of TACOPlug with exiting applications. 

\begin{figure}
     \centering
      \begin{subfigure}[b]{\linewidth}
         \centering
         \begin{equation*}
              y(i) = A(i,j) * x (j)
         \end{equation*}
         \caption{Tensor Index Notation}
         \label{fig:tensor_index_notation}
     \end{subfigure}
     
      \begin{subfigure}[b]{\linewidth}
         \centering
         \begin{lstlisting}
taco  "y(i) = A(i,j) * x (j)"\\
    -f=A:ds:0,1 -f=x:d -f=y:d
         \end{lstlisting}
         \caption{Taco Command Line}
         \label{fig:tensor_command_line}
     \end{subfigure}
    
     \begin{subfigure}[b]{\linewidth}
         \centering
         \begin{lstlisting}
Format csr({Dense,Sparse});
Format dv({Dense});
Tensor<double> A;
Tensor<double> x({A.getDimension(1)}, dv);
Tensor<double> y({A.getDimension(0)}, dv);
IndexVar i,j;
y(i) = A(i,j) * x (j);
y.compile();
y.assemble();
y.compute();
         \end{lstlisting}
         \caption{Taco DSL}
         \label{fig:taco_dsl}
     \end{subfigure}
     \hfill
     
        \caption{Sparse Matrix Vector Multiplication }
        \label{fig:three graphs}
\end{figure}

\section{Quantum Programming in C\texttt{++}}
\label{sec:qcor}

 Quantum computation has emerged as a potential avenue for the continued scalability of high-performance scientific computing. Current quantum processing units (QPUs) developed by vendors such as IBM, Rigetti, and Google are available for experimentation over the cloud, and a number of interesting toy-model simulations have been demonstrated. These novel processing units provide a small number of qubits, limited connectivity, and are subject to decoherence and other noise sources that limit the overall utility for large scale simulations. However, as quantum architectures continue to advance, one could imagine the future integration of robust QPUs with classical heterogeneous compute resources via a typical co-processor or accelerated computing model. There is therefore a need for research into, and the development of, quantum-classical software infrastructures, languages, and compilers enabling hybrid quantum-classical algorithmic expression. 

C\texttt{++} has proven ubiquitous in the high-performance computing research space for its performance, portability, and multi-paradigm expressibility. Future CPU-QPU models will require tight integration and fast-feedback capabilities enabling quantum sub-program execution based on qubit measurement results. C\texttt{++} is well-positioned to provide the necessary language and programming model that will enable this tight integration. We therefore seek to extend C\texttt{++} with support for quantum programming, enabling integration of standard C\texttt{++} language utilities with common, popular methods for the expression of quantum code at varying levels of abstraction. 
\begin{figure}[b!]
\centering  
\includegraphics[width=.45\textwidth]{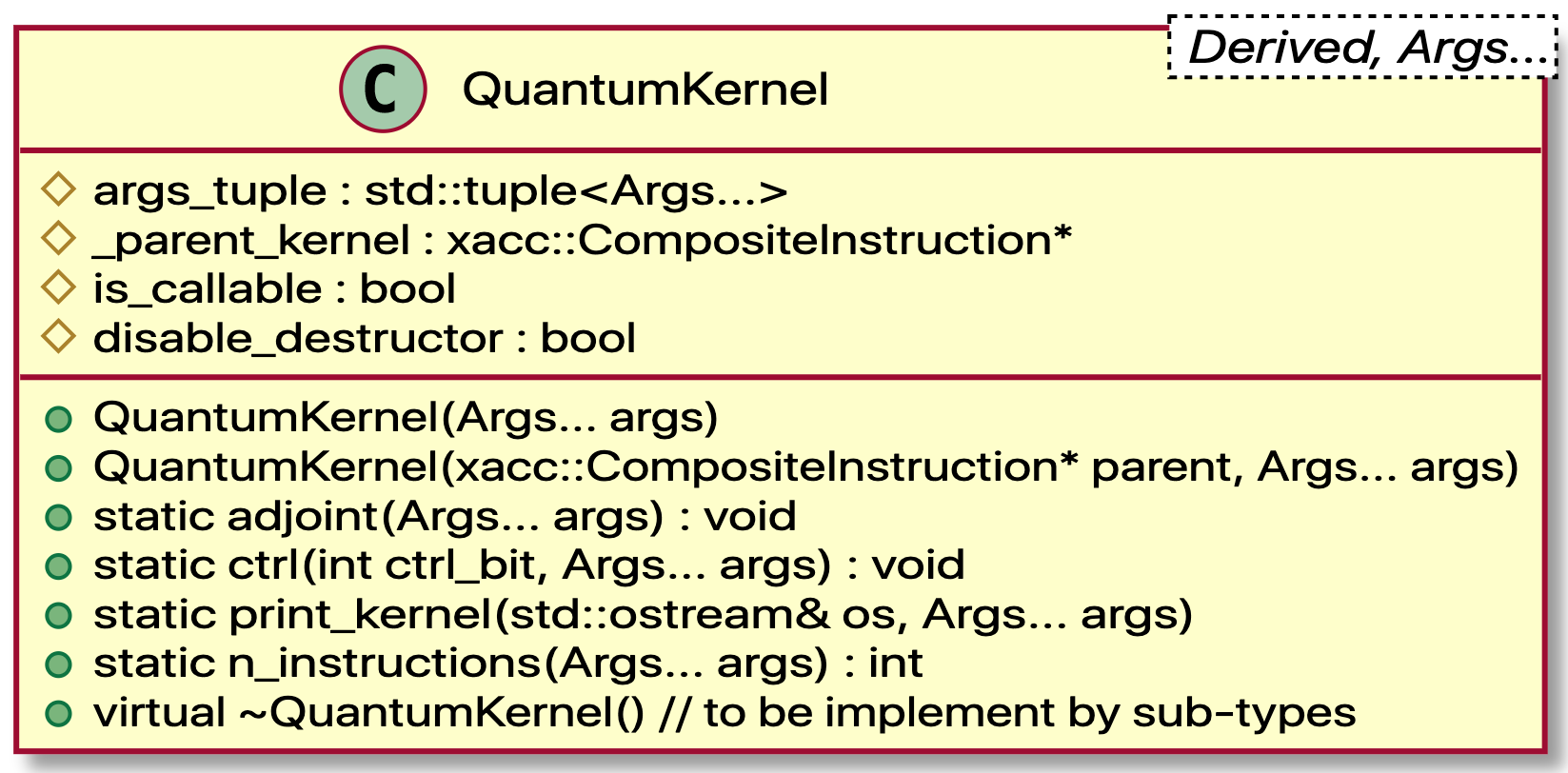}
\caption{The class diagram for the \texttt{QuantumKernel} template class.}
\label{fig:qkernel_uml}
\end{figure}
\begin{figure}[t!] 
  \lstset {language=C++}
\begin{lstlisting}
// Programmer code                            (A.)
[[clang::syntax(quantum)]] 
    void ansatz(qreg q, double x) {
  X(q[0]);
  Ry(q[1], x);                     
  CX(q[1], q[0]);
}
[[clang::syntax(quantum)]] 
    void do_nothing(qreg q, double x) {
  ansatz(q, x);
  ansatz::adjoint(q,x);
}
[[clang::syntax(quantum)]] 
    void use_ctrl(qreg q, double x) {
  // qreg of size 3
  ansatz::ctrl(q[2], q, x);
}
---------- ansatz translates to ------------------
// SyntaxHandler-generated code for ansatz() (B.)
void ansatz(qreg q, double x) {
  void internal_ansatz_call(qreg, double);
  internal_ansatz_call(q, x);
}                                  
class ansatz : 
    public QuantumKernel<ansatz, qreg, double> {
public:
  ansatz(qreg q, double x) : 
    QuantumKernel<ansatz, qreg, double>(q,x) {}
  virtual ~ansatz() {
   auto [q,x] = args_tuple;
   // -------------------------------------------
   // Generated from Token Analysis
   auto provider = xacc::getIRProvider();
   auto i0 = provider->createInstruction("X",{0});
   auto i1 = 
       provider->createInstruction("Ry",{1}, {x});
   auto i2 = provider->createInstruction("CX",{1,0});
   _parent_kernel->addInstructions({i0,i1,i2});
   // -------------------------------------------
   auto qpu = xacc::getAccelerator("ibm:ibmq_viqo");
   qpu->execute(q, _parent_kernel);
  }
}
void internal_ansatz_call(qreg q, double x) {
 class ansatz temporary_instance(q, x);
}
----------- used as function call ----------------
// Using the quantum kernel                   (C.)
int main() {
  // allocate 2 qubits
  auto q = qalloc(2);              
  // execute the quantum code
  ansatz(q, 1.57079632679);
  // show results
  q.print();
  return 0;
}
\end{lstlisting}
\caption{Code snippet demonstrating quantum kernel programming (A.), its representation after SyntaxHandler processing (B.), and how programmers might invoke the quantum kernel (C.).}
\label{fig:sh_code}
\end{figure}

Here we describe how we leverage the new Clang SyntaxHandler plugin to process functions written in quantum domain specific languages and translate them to appropriate C++ functions composed of valid API calls. Specifically, we leverage the XACC quantum programming framework which provides a collection of C++ libraries enabling the standard quantum programming, compilation, and execution workflow \cite{xacc}. XACC enables a robust API for quantum circuit composition, compilation, and hardware-agnostic execution, with support for QPUs provided by IBM, Rigetti, D-Wave, IonQ, and Honeywell. Ultimately the goal of our SyntaxHandler is to translate general quantum functions in a way that leverages XACC but also provides convenient, automatically-generated related circuits. This model should promote general quantum circuit expressibility, kernel composability, and higher-levels of quantum code abstraction. 



Figure \ref{fig:qkernel_uml} provides the QuantumKernel class description, the underlying object model that we leverage to represent quantum kernel functions. Our custom SyntaxHandler translates these functions into sub-type definitions of this class. QuantumKernel is templated on the derived type and the kernel function argument types. It exposes constructors that take kernel function arguments as input, and stores them in a protected \texttt{args\_tuple} class attribute. QuantumKernels keep reference to an XACC CompositeInstruction (intermediate representation of a compiled quantum circuit) and the goal of sub-types is to construct this \texttt{\_parent\_kernel} CompositeInstruction dependent on the original quantum kernel definition (before SyntaxHandler processing). QuantumKernel exposes static public methods for automatically generating related circuits for common use cases such as controlled unitary operation and adjoint, or reversal, of the current quantum code. The goal of sub-types is to implement the class destructor such that the internal CompositeInstruction is constructed (representing the quantum code) and to submit or execute that CompositeInstruction on the targeted quantum backend. Note that this process makes use of the input \texttt{args\_tuple} for this quantum circuit generation. This feature enables quantum kernel invocation via temporary class instance construction followed by immediate destruction, but also allows internal development libraries to leverage the full class API for more complex use cases. 

Figure \ref{fig:sh_code} demonstrates how users can program and use quantum kernel functions and how the SyntaxHandler can translate these invalid function definitions into valid C\texttt{++} code. The workflow put forward by our \texttt{SyntaxHandler::GetReplacement()} begins by converting the vector of Clang Tokens into individual, constituent statements (e.g. in the simplest case, combine Tokens to form a string, stopping at all semicolons). Each of these statements is analyzed to produce an equivalent XACC instruction creation API call that leverages the \texttt{xacc::IRProvider} factory pattern. The SyntaxHandler then sets up the new code to add these instructions to the \texttt{QuantumKernel::\_parent\_kernel} and submit it to the backend \texttt{xacc::Accelerator} for execution (for more information on these XACC data types see \cite{xacc}). This new code string made up of XACC API calls is then added to the destructor call of a newly described QuantumKernel sub-type generated by the SyntaxHandler. This sub-type is of the same name as the original kernel function, and the correct template parameters for the QuantumKernel are configured based on the original function arguments. The original function is re-written to contain a forward declaration to an internal function call that is then immediately called. This function is defined after the class is defined to make use of the temporary instance construction-destruction model of the quantum kernel construction and execution. Since the quantum kernel function is actually a class sub-type of the QuantumKernel after processing of the SyntaxHandler, \texttt{ctrl} and \texttt{adjoint} methods are also available for use within further quantum kernel function definitions (see the second and third functions in section (A.) of the code snippet in Figure \ref{fig:sh_code}). 

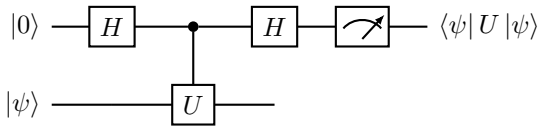
\begin{figure}[b!]
\begin{quantikz}
\lstick{$\ket{0}$} & \gate{H} & \ctrl{1} & \gate{H} & \meter{} & \rstick{$\bra{\psi} U \ket{\psi} $} \qw \\
\lstick{$\ket{\psi}$} & \ghost{H}\qw & \gate{U} & \qw
\end{quantikz}
\caption{Quantum circuit implementing the Hadamard test.}
\label{fig:htest_circ}
\end{figure}

\begin{figure}[t!] 
  \lstset {language=C++}
\begin{lstlisting}
// Hadamard Test, Compute <psi | U | psi>, 
// Here U = X (pauli x matrix)
__qpu__ void x_gate(qreg q) { X(q[1]); }
__qpu__ void hadamard_test_x(qreg q) {
  // Create the superposition on the first qubit
  H(q[0]);

  // Apply U = X on second qubit
  x_gate(q);

  // apply ctrl-U = C-X, 
  // use QuantumKernel::ctrl
  x_gate::ctrl(q[0], q);

  // add the last hadamard
  H(q[0]);

  // measure
  Measure(q[0]);
}
int main() {
  // allocate 2 qubits
  auto q = qalloc(2);
  // run the hadamard test
  hadamard_test_x(q);
  // Get the number of times 1 and 0 were seen
  auto count1 = q.counts().find("1")->second;
  auto count0 = q.counts().find("0")->second;
  // Compute <psi|X|psi> 
  std::cout << "<X> = " <<
            << std::fabs((count1 - count0) 
            / (double)(count1 + count0)) << "\n";
}
// Compile with 
// clang++ -fplugin=/path/to/quantum-syntax.so ...
\end{lstlisting}
\caption{Code snippet demonstrating a simple Hadamard test. This example demonstrate kernel composability and the utility of the QuantumKernel auto-generated \texttt{ctrl} method.}
\label{fig:htest}
\end{figure}
As a final example of the programmability of a typical benchmark problem in quantum computing leveraging a programming model that uses this novel Clang SyntaxHandler, we present an example known as the Hadamard test. The goal of this test is, given some unitary matrix $U$ and some initial quantum state $\ket{psi}$, compute the expected value of $U$ with respect to that state, $\bra{\psi} U \ket{\psi}$. The quantum circuit for achieving this is shown in Figure \ref{fig:htest_circ}, where one prepares a superposition state on the first qubit (via the application of a Hadamard, H), then prepares a second qubit register into the state $\ket{\psi}$, operates a controlled-$U$ application from the first qubit onto that second qubit register, and finally applies a second Hadamard followed by measurement. The expected value can be computed as the difference between the probability of seeing the $\ket{1}$ and $\ket{0}$ states. We present a program in Figure \ref{fig:htest} that implements this problem for $U=\sigma_x$ and $\ket{\psi}=\ket{1}$. We define one quantum kernel function that represents that unitary $U$ application, \texttt{x\_gate}, and another that runs the Hadamard test itself, \texttt{hadamard\_test\_x}. As one can see, this kernel function starts by applying the first Hadamard gate on the first qubit. We then create $\ket{\psi}=U\ket{0}$ by calling to the \texttt{x\_gate} kernel. Next we leverage the auto-generated \texttt{ctrl} method coming from the QuantumKernel super-class to apply the controlled unitary operation. Finally we add another Hadamard and measure the first qubit. In \texttt{main} we allocate 2 qubits, run the Hadamard test, and compute $\bra{\psi} U \ket{\psi}$, printing it to \texttt{std::cout}.

\section{TAProL-Plug: Embedding Tensor Network Theory in C++}
\label{sec:taprol}

TAProL (Tensor Algebra Programming Language) is a concise mathematical-notation DSL for expressing numerical tensor algebra algorithms dealing with tensor networks (it is part of the ExaTN library~\cite{exatn}). It supports basic tensor operations, like tensor scaling, tensor addition, tensor product, tensor contraction, as well as higher-level tensor operations, like tensor decompositions. The latter are formalized via the tensor network theory, where a tensor network approximates a given higher-order tensor as a specifically structured contraction of lower-order tensors. The specific form of the tensor contraction used, that is, the topology of the tensor network, determines its class, with the most widely used classes being the tensor train (TT) or, equivalently, matrix product state (MPS)~\cite{mps}, projected entangled pair state (PEPS)~\cite{peps}, tensor tree network (TTN)~\cite{ttns}, or, equivalently, the hierarchical Tucker decomposition, and multiple entanglement renormalization ansatz (MERA)~\cite{mera}, to name a few. Low-rank tensor decompositions and factorizations based on the tensor networks have been successfully utilized for approximating quantum many-body wavefunctions in condensed matter physics~\cite{mera}, quantum chemistry~\cite{ttns_chan} and quantum computing~\cite{mps_tnqvm}. Recently they have also been used for the tensor completion problem~\cite{tensor_completion} as well as neural network layer regularization in machine learning~\cite{dnn_tn_regular}.

\begin{figure}[t!]
\centering
\begin{lstlisting}
[[clang::syntax(taprol)]]
void test(std::vector<std::complex<double>>& t2_data,
          std::shared_ptr<talsh::Tensor> talsh_t2,
          std::shared_ptr<talsh::Tensor> talsh_x2,
          double& norm_x2) {

 //Declaring the TAProL entry point:
 entry: main;
 //Opening a TAProL scope (optional):
 scope main group(tensor_workload);
  //Declaring linear spaces of some dimension:
  space(complex): space0 = [0:255], space1 = [0:511];
  //Declaring subspaces of declared linear spaces:
  subspace(space0): s0 = [0:127], s1 = [128:255];
  subspace(space1): r0 = [0:283], r1 = [284:511];
  //Associating index labels with declared subspaces:
  index(s0): i, j, k, l;
  index(r0): a, b, c, d;
  //Initializing a tensor to zero:
  Z2(a, b, i, j) = {0.0, 0.0};
  //Initializing a tensor by a registered functor:
  H2(a, i, b, j) = method("ComputeHamiltonian");
  //Initializing a tensor by external C++ data:
  T2(a, b, i, j) = t2_data;
  //Contracting two tensors:
  Z2(a, b, i, j) += H2(a, k, c, i)
                  * T2(b, c, k, j);
  //Contracting three tensors (tensor network):
  Z2(a, b, i, j) += H2(c, k, d, l)
                  * T2(c, d, i, j)
                  * T2(a, b, k, l);
  //Scaling a tensor by a scalar:
  Z2(a, b, i, j) *= 0.25;
  //Adding a tensor to another tensor:
  T2(a, b, i, j) += Z2(a, b, i, j);
  //Decomposing tensor T2 into two factors:
  A(a, b, c) = {0.0, 0.0};
  B(i, j, c) = {0.0, 0.0};
  A(a, b, c) * B(i, j, c) = T2(a, b, i, j);
  //Exporting tensor T2 back to C++:
  talsh_t2 = T2;
  //Computing the 2-norm of tensor Z2:
  X2() = {0.0, 0.0};
  X2() += Z2+(a, b, i, j)
        * Z2(a, b, i, j);
  norm_x2 = norm1(X2);
  //Exporting scalar X2 back to C++:
  talsh_x2 = X2;
  //Saving scalar X2:
  save X2: tag("Z2_norm2");
  //Destroying tensors X2 and Z2:
  ~X2;
  ~Z2;
  //Destroying other tensors (alternative):
  destroy A, B, T2, H2;
 end scope main;
}
\end{lstlisting}
\caption{TAProL code example}
\label{fig:taprolplug}
\end{figure}
         
The TAProL code consists of a mix of declarative and executive statements (see Fig. \ref{fig:taprolplug}). Declarative statements are used to declare linear spaces, their subspaces, and indices associated with those subspaces. These spaces/subspaces are used for defining tensors (each tensor dimension is associated with a declared subspace). The executive statements define either simple or composite tensor operations. Simple tensor operations are tensor initialization, tensor scaling, tensor addition, tensor product, binary tensor contraction, etc. Tensor initialization has a number of variants. In particular, one can initialize a tensor from externally provided data stored in a C++ std::vector container. Additionally, one can also initialize or transform a tensor using a user-defined C++ functor, thus enabling custom unary operations on TAProL tensors written in C++. Composite tensor operations include (a) arbitrary tensor contractions of three or more tensors (contraction of a tensor network), and (b) tensor decompositions, which are the opposites of the tensor contractions (see Fig. \ref{fig:taprolplug}). To export a TAProL tensor back to the C++ realm, one can use a pre-defined C++ class talsh::Tensor provided by the ExaTN library. For further convenience, TAProL statements can be organized into scopes (TAProL analogs of routines) inside a C++ function.

During the compilation, the TAProL clang syntax plugin translates the TAProL code into the calls to the ExaTN library C++ API. Simple tensor operations are scheduled for asynchronous execution via a dynamic task graph during program execution. Composite tensor operations are first decomposed into simple tensor operations which are subsequently added into the task graph during program execution as well. The parallel runtime of the ExaTN library attempts to automatically decompose and distribute tensor operations among all computing devices across a given set of compute nodes. The runtime also automatically tracks all data dependencies, thus allowing the user to write serial programs that will be transparently parallelized by the ExaTN library runtime. This enables the execution of tensor algebra workloads on a computer of any scale, from laptops to leadership HPC systems.

\section{Conclusion}
\label{sec:con}

The syntax plugin infrastructure allows for the efficient, productive integration of DSLs into C++ code bases. Several examples were provided, ranging from quantum computing to sparse and multi-dimensional linear algebra. Moreover, this work demonstrates the utility of extending C++ compilers with powerful plugin interfaces in order to enable novel use cases and provide more-pleasant development environments. In the future, we expect additional DSLs, or maybe even other general-purpose languages (e.g., Python), to be embedded into C++ using syntax plugins. As the technology becomes more wide spread, debuggers, profilers, and other programming-environment tools will need to be enhanced to understand syntax plugins. The syntax-plugin interface itself might grow to add more capabilities. For example, it might be demonstrated that injecting DSL code in more places than just function bodies is useful. Finally, support syntax plugins should be added to the mainline Clang project, and hopefully, similar interfaces will also be added to other C++ implementations. The overall result will be that programmers will be able to increase their productivity by using DSLs without some of the trade offs that currently entails.

\section*{Acknowledgment}
This work has been supported by the US Department of Energy (DOE) Office of Science Advanced Scientific Computing Research (ASCR) Quantum Computing Application Teams (QCAT), Quantum Algorithms Team (QAT), and Accelerated Research in Quantum Computing (ARQC). This research was also supported by the Exascale Computing Project (17-SC-20-SC), a collaborative effort of the U.S. Department of Energy Office of Science and the National Nuclear Security Administration. ORNL is managed by UT-Battelle, LLC, for the US Department of Energy under contract no. DE-AC05-00OR22725. This research used resources of the Argonne Leadership Computing Facility, which is a DOE Office of Science User Facility supported under Contract DE-AC02-06CH11357. This research used resources of the Oak Ridge Leadership Computing Facility, which is a DOE Office of Science User Facility supported under Contract DE-AC05-00OR22725. We would also like to acknowledge the Laboratory Directed Research and Development (LDRD) funding from the Oak Ridge National Laboratory (award 9463).




\bibliographystyle{IEEEtran}
\bibliography{bibliography}

\end{document}